\documentclass[a4paper,11pt]{article}
\pdfoutput=1
\usepackage{amsmath,amssymb,graphicx,todonotes,xspace,cite,hyperref,slashed,subcaption,float,tabularx}
\usepackage{booktabs}
\usepackage{listings}
\usepackage[section]{placeins}
\hypersetup{
  pdftitle={HEJ 2.1},
  colorlinks,
  urlcolor={blue},
  linkcolor={blue},
}

\graphicspath{{figures/}{build/figures/}}

\lstdefinelanguage{sherpa}{%
  morekeywords={run,processes,selector},
  sensitive=false,
  morecomment=[l]{\#},
}

\lstdefinelanguage{yaml}{%
  morekeywords={},
  sensitive=false,
  morecomment=[l]{\#},
}

\newcommand{\HEJ}{{\tt HEJ}\xspace}
\newcommand{\HEJFOG}{{\tt HEJFOG}\xspace}

\newcommand{\HIGHEJ}{\emph{High Energy Jets}\xspace}
\newcommand{\tinyspace}{\mkern 1mu}

\title{\begin{normalsize}
\begin{flushright}
DCPT/21/86, \\DESY-21-174,\\ IPPP/21/43,\\MCNET-21-14,\\ SAGEX-21-33
\end{flushright}
\end{normalsize}
\vspace*{2cm}\HEJ 2.1: High-energy Resummation with Vector Bosons and Next-to-Leading Logarithms}

\author{%
  \begin{minipage}{1.0\linewidth}
    \begin{center}
  Jeppe R.~Andersen$^{a}$,
  James Black$^{a}$,
  Helen Brooks$^{b}$,
  Bertrand Duclou\'e$^{c}$,
  Marian Heil$^{a}$,
  Andreas Maier$^{d}$, %
  Jennifer M.~Smillie$^{c}$
\end{center}
\end{minipage}
}
\date{%
$^{a}$ Institute for Particle Physics Phenomenology,\\ University of Durham,
Durham, DH1 3LE, UK\\%
$^{b}$ School of Physics and Astronomy, Monash University, Clayton, VIC 3800, Australia\\
$^{c}$ Higgs Centre for Theoretical Physics, University of Edinburgh,\\
Peter Guthrie Tait Road, Edinburgh, EH9 3FD, UK\\%
$^{d}$ Deutsches Elektronen-Synchrotron DESY,\\ Platanenallee 6,
  15738 Zeuthen, Germany}

\begin{document}

\maketitle

\abstract{%
  We present version 2.1 of the \HIGHEJ (\HEJ) event generator for
  hadron colliders. \HEJ is a Monte Carlo generator for processes at
  high energies with multiple well-separated jets in the final state.
  To achieve accurate predictions, conventional fixed-order perturbative
  QCD is supplemented with an all-order resummation of large high-energy
  logarithms.  The new version 2.1 now supports processes with
  final-state leptons originating from a charged or neutral vector boson
  together with multiple jets, in addition to processes available in
  earlier versions. Furthermore, the all-order resummation is extended
  to include an additional gauge-invariant class of subdominant
  logarithmic corrections. \HEJ 2.1 can be obtained from
  \url{https://hej.hepforge.org}.
}

\newpage
\tableofcontents

\section{Introduction}
\label{sec:intro}

\HIGHEJ (\HEJ) is a Monte Carlo event generator for processes
involving two or more jets. It implements the eponymous formalism for
the all-order summation of high-energy logarithms in
$\tfrac{\hat{s}}{p_t^2}$ developed
in~\cite{Andersen:2009nu,Andersen:2009he,Andersen:2011hs}. This
summation is necessary to obtain a good description of events with
large invariant masses or large rapidity separations between the
outgoing particles. Extensive reviews of the formalism can be found
in~\cite{Andersen:2017kfc,Andersen:2020yax}.

\HEJ has been validated against data in numerous studies on pure
multijet
production~\cite{ATLAS:2011yyh,CMS:2012rfo,CMS:2012xfg,ATLAS:2014lzu},
production of a leptonically decaying W boson together with two or
more jets~\cite{Andersen:2012gk,D0:2013gro,ATLAS:2014fjg}, and the
production of multiple jets together with two charged leptons, created
from an intermediate photon or Z boson~\cite{Andersen:2016vkp}. A
further particularly interesting channel is the gluon-fusion
production of a Higgs boson together with two or more jets. It
constitutes the dominant background in measurements of Higgs boson
production in weak boson fusion (VBF), and the application of typical
VBF cuts projects out a region of phase space where the summation of
high-energy logarithms yields significant
corrections~\cite{Andersen:2008gc}. For this process, as well as for
pure multijet production, the original leading-logarithmic (LL)
summation was supplemented with a numerically important
gauge-invariant subset of the next-to-leading-logarithmic (NLL)
corrections to achieve a better description away from the asymptotic
high-energy limit~\cite{Andersen:2017kfc}.

A complete redesign allowed to extend the matching between resummation
and fixed-order predictions to the highest
multiplicities~\cite{Andersen:2018tnm}, taking into account at the
same time quark mass corrections in the gluon-fusion production of a
Higgs boson together with jets~\cite{Andersen:2018kjg}. This revised
matching procedure, initially implemented for pure multijet and Higgs
boson plus jets production, is the foundation of \HEJ2~\cite{Andersen:2019yzo}.

Here, we present \HEJ~2.1. This new version elevates the fixed-order
matching for the aforementioned processes involving an intermediate W
or Z boson or photon to the superior \HEJ2 procedure. Furthermore, it
includes a new subset of NLL corrections for pure multijet production
and the production of a W boson together with jets. With these
additions, \HEJ~2.1 supersedes all earlier \HEJ versions. We briefly
describe the addition of the processes involving intermediate vector
bosons and the new NLL corrections associated with the emission of
additional quark-antiquark pairs in section~\ref{sec:features}. In
section~\ref{sec:application}, we give an explicit example
demonstrating how \HEJ~2.1 can be used to supplement leading-order
descriptions of the production of two leptons with jets with
high-energy resummation. We conclude in section~\ref{sec:conclusions}.


\section{Improvements over previous versions}
\label{sec:features}

The most significant improvements over previous versions of \HEJ are
the addition of high-energy resummation for the production of two
leptons together with two or more jets and the inclusion of NLL
corrections involving the production of an additional quark-antiquark
pair. In the following, we briefly summarise how these new processes
are described in the \HIGHEJ formalism. A more detailed description of
both the underlying formalism and the improvements discussed here is given
in~\cite{Andersen:2020yax}. The changes are summarised in table~\ref{tab:summary}.
\begin{table}[htb]
  \centering
  \begin{tabular}{lccc}
    \toprule
    Process & LL & NLL unordered gluon & NLL quark-antiquark \\
    \midrule
    $\geq$2 jets & \HEJ 2.0 & \HEJ 2.0 & \HEJ 2.1\\
    H + $\geq$2 jets & \HEJ 2.0 & \HEJ 2.0 & ---\\
    W + $\geq$2 jets & \HEJ 2.1 & \HEJ 2.1 & \HEJ 2.1\\
    Z/$\gamma$ + $\geq$2 jets & \HEJ 2.1 & \HEJ 2.1 & ---\\
    \bottomrule
  \end{tabular}
  \caption{Implemented processes and higher-order logarithmic corrections in \HEJ}
  \label{tab:summary}
\end{table}

\subsection{Leading-logarithmic resummation in \HEJ2}
\label{sec:HEJ2_res}

In \HEJ2, fixed-order predictions are supplemented with high-energy
resummation in the following way. First, a number of leading-order
events are generated. Then, for each event, it is determined whether
the corresponding matrix element contributes at a logarithmic accuracy
that is included in the current state-of-the art \HEJ
resummation. While \HEJ currently includes partial NLL resummation, we
restrict the following discussion to LL resummation for the sake
of clarity.

\begin{figure}[htb]
  \centering
  \begin{tabularx}{\linewidth}{cXcXcXl}
  \includegraphics[height=0.15\linewidth]{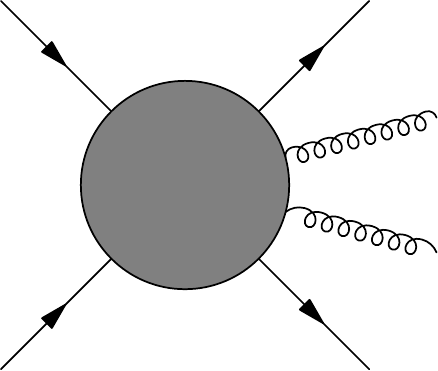}&&
  \includegraphics[height=0.15\linewidth]{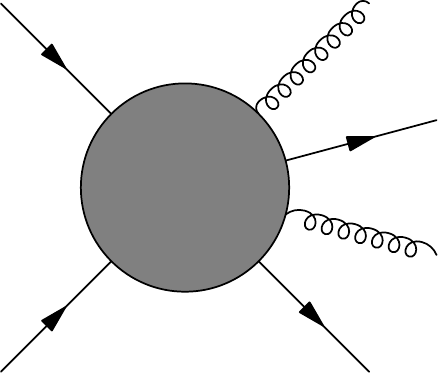}&&
  \includegraphics[height=0.15\linewidth]{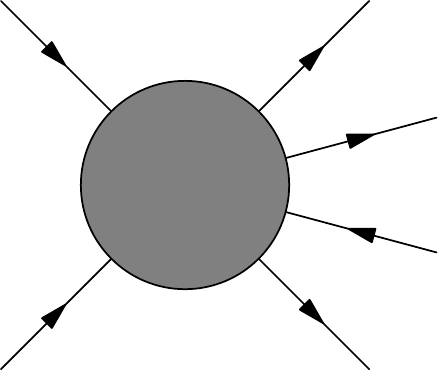}&&
  \includegraphics[height=0.15\linewidth]{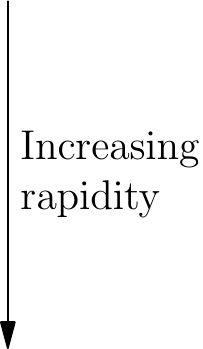}
\end{tabularx}
  \caption{Identification of configurations with leading-logarithmic
contributions. The leftmost configuration allows the maximum of three
$t$-channel gluon exchanges and therefore contributes at LL
accuracy. The remaining configurations permit at most two $t$-channel
gluons. They are suppressed in the high-energy limit.}
  \label{fig:conf}
\end{figure}

To identify event configurations contributing at LL accuracy, we order
incoming and outgoing particles according to rapidity, see
figure~\ref{fig:conf}. We then draw a leading-order auxiliary diagram
with as many $t$-channel gluons as possible. It should be stressed
that this diagram only serves to determine the logarithmic order and
is never used to compute the actual matrix element. A configuration
contributes at LL accuracy iff the number of $t$-channel gluons is
maximal, i.e. if no other rapidity ordering of an equivalent set of
final-state particles allows a diagram with a larger number of
$t$-channel gluon exchanges. By ``equivalent'' we mean that we do not
distinguish between parton flavours, such that all final states shown
in figure~\ref{fig:conf} are treated on the same footing.

If the matrix element of the currently considered event contributes at
a logarithmic order that is covered by the resummation, we replace the
original event with a number of newly generated events. These new
events add further gluon emissions, while preserving the rapidities of
the original jets and non-parton particles. Details of the procedure
are given in~\cite{Andersen:2018tnm}.  In order to obtain an all-order
resummed prediction, while at the same time retaining leading-order
accuracy, the newly generated events are reweighted by a factor
\begin{equation}
  \label{eq:rew}
  w_{\text{res}} = \frac{\overline{|{\cal M}^{\text{\HEJ}}|}^2}{\overline{|{\cal M}^{\text{\HEJ}}_{\text{LO}}|}^2},
\end{equation}
where ${\cal M}^{\text{\HEJ}}$ is the all-order resummed matrix
element in the high-energy limit and ${\cal
M}^{\text{\HEJ}}_{\text{LO}}$ its leading-order truncation. The bar
denotes the sum (average) over outgoing (incoming) helicities and
colours.

The \HEJ matrix elements are the only process-specific ingredient in
the resummation procedure. In the following, we recall their general
structure and give results for the new processes included in \HEJ~2.1.

\subsection{Leading-logarithmic matrix elements}
\label{sec:ME_LL}

We are interested in the LL \HEJ matrix elements for processes $f_a
f_b \to Xf_{a'}\cdot ng \cdot f_{b'}$, where $X$ denotes any number of
additional colourless outgoing particles. With the exception of $X$,
particles are ordered according to rapidity. $f_a$ is therefore the
incoming parton in the backward direction, with flavour $a$, and $f_b$
the parton in the forward direction. The most backward outgoing parton
is $f_{a'}$ with rapidity $y_1$, followed by $n$ gluons with
rapidities $y_2 < \dots < y_{n+1}$ and the most forward outgoing
parton $f_{b'}$ with rapidity $y_{n+2}$.

In the absence of $X$, the outgoing flavours match the incoming ones,
i.e. $a' = a, b' = b$. This is still the case if a charged
lepton-antilepton pair $X = l\bar{l}$ is produced. However, for this
process we require a virtual photon or Z boson, which implies that at
least one of $f_a$ and $f_b$ has to be a quark or antiquark. When
producing a pair of charged lepton and a neutrino, $X = l \bar{\nu}_l$
or $X = \bar{l} \nu_l$, a quark or antiquark couples to a virtual W
boson. This means that the respective flavour is changed, so that
either $a' \neq a$ or $b' \neq b$.

\subsubsection{Matrix element without interference}
\label{sec:ME_no_interference}

\begin{figure}[htb]
  \centering
  \includegraphics{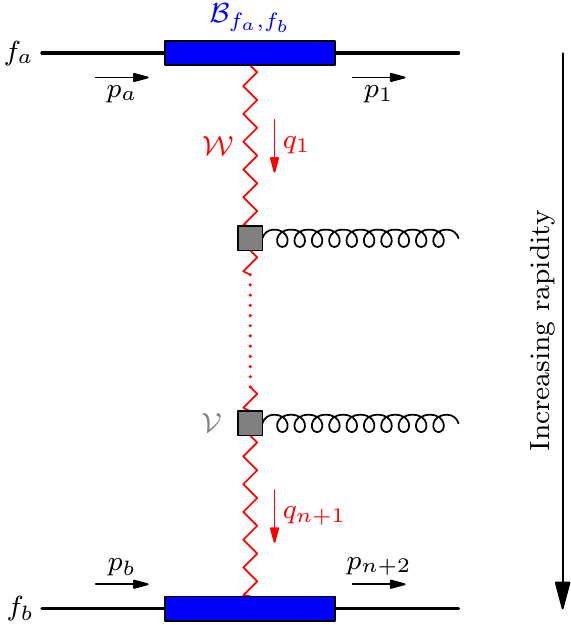}
  \caption{Structure of the LL \HEJ matrix element in the absence of $X$.}
  \label{fig:ME_LL}
\end{figure}

The square of the LL \HEJ matrix element in the absence of
interference has the following structure, illustrated also in figure~\ref{fig:ME_LL}:
\begin{equation}
  \label{eq:ME_LL}
  \begin{split}
\overline{|{\cal M}^{\text{\HEJ}}_{f_a f_b \to Xf_{a'}\cdot ng \cdot f_{b'}}|} ={}
&\textcolor{blue}{{\cal B}_{f_a,f_b,X}(p_a, p_b, p_1, p_{n+2}, \{p\}_X)}\\
&\cdot \prod_{i=1}^n \textcolor{gray}{{\cal V}(p_a, p_b, p_1, p_{n+2}, q_i, q_{i+1})}\\
&\cdot \prod_{i=1}^{n+1} \textcolor{red}{{\cal W}(q_j,y_j,y_{j+1})}
    \end{split}
  \end{equation}
$p_a$ and $p_b$ are the momenta of $f_a$ and $f_b$, $p_i$ with $1 \leq
i \leq n+2$ the outgoing momenta ordered by ascending rapidity, and
$q_j$ the $t$-channel momenta related by $q_j = q_{j-1} - p_j$. The
first $t$-channel momentum $q_1$ is process-dependent; if there are no
additional particles $X$ it is given by $q_1 = p_a - p_1$.  $\{p\}_X$
represents the local momenta associated with the production of $X$.

${\cal B}_{f_a,f_b,X}$ is the square of the Born-level matrix element for the
process without any additional gluon emissions, $n=0$. ${\cal V}$
describes the real gluon emission corrections to this process, whereas
${\cal W}$ accounts for the virtual corrections. Both are
process-independent and given by
\begin{align}
  \label{V}
  {\cal V}(p_a, p_b, p_1, p_{n+2}, q_i, q_{i+1}) ={}& -\frac{C_A}{q_i^2q_{i+1}^2} V^\mu(q_i, q_{i+1}) V_\mu(q_i, q_{i+1}),\\
  \label{W}
  {\cal W}(q_j,y_j,y_{j+1}) ={}& \exp[\omega^0(q_{j\perp}) (y_{j+1}- y_j)],
\end{align}
where $q_{j\perp}$ is the component of the $t$-channel momentum $q_j$
that is perpendicular to the beam axis. Using a short-hand notation,
we have suppressed the dependence of the Lipatov vertex function
$V^\mu$ on the incoming and outgoing momenta:
\begin{equation}
  \label{eq:Lipatov_shorthand}
  V^\mu(q_i, q_{i+1}) = V^\mu(p_a, p_b, p_1, p_{n+2}, q_i, q_{i+1}).
\end{equation}
Explicit expressions for $V^\mu$ and the regularised Regge trajectory
$\omega^0$ are derived in~\cite{Andersen:2009nu,Lipatov:1976zz}.

The only process-dependent part is the Born function ${\cal
B}_{f_a,f_b,X}$.  It is derived via matching to full QCD, i.e.~by
requiring that ${\cal B}$ is equivalent to the corresponding QCD
amplitude in the high-energy limit. In the \HIGHEJ formalism, exact
gauge invariance and superior numerical agreement with full QCD over
the whole phase space are achieved by only neglecting a
gauge-invariant subset of the terms that are supressed in the
high-energy limit. Specifically, only those terms that would break the
following $t$-channel factorised form are discarded.
\begin{equation}
  \label{eq:B}
  {\cal B}_{f_a,f_b,X}(p_a, p_b, p_1, p_{n+2}, \{p\}_X) = (g_s^2)^2 \frac{K_{f_a} K_{f_b}}{4 (N_C^2 - 1)} \|S_{f_a f_b \to Xf_{a'}\dots f_{b'}}\|^2 \frac{1}{q_1^2 q_{n+1}^2}.
\end{equation}
$K_{f_a}$ and $K_{f_b}$ are colour factors depending on the incoming
partons. For an (anti-)quark $i$ the corresponding factor is simply
$K_{f_i} = C_F$. Gluons lead to more complex factors $K_g$ involving
also their kinematics~\cite{Andersen:2009he}. Finally, $S_{f_a f_b \to
Xf_{a'}\dots f_{b'}}$ is the contraction of two (generalised)
currents, with double bars indicating the sum over helicities. In pure
QCD, $X$ is absent and the current contraction is simply
\begin{equation}
  \label{eq:S_QCD}
  \| S_{f_a f_b \to f_a \dots f_b}\|^2 \equiv \| j^a \cdot j^b\|^2 = \sum_{h_a, h_b}|j^{\mu,h_a}(p_1, p_a) j_\mu^{h_b}(p_{n+2}, p_b)|^2,
\end{equation}
where $j_\mu^h$ is the current
\begin{equation}
  \label{eq:j}
  j_\mu^h(p,q) = \bar{u}^h(p) \gamma_\mu u^h(q)
\end{equation}
for helicity $h$.

A pair consisting of a lepton\footnote{We do not distinguish between
charged leptons and neutrinos here.} with momentum $p_\ell$ and an
antilepton with momentum $p_{\bar{\ell}}$ is produced via a virtual
vector boson $V \in \{W^+,W^-,Z,\gamma\}$ coupling to either of the
partons $f_a$ or $f_b$. We can account for this process by modifying
the corresponding current. Assuming without loss of generality a
coupling to $f_a$, we obtain
\begin{equation}
  \label{eq:S_V}
  \| S_{f_a f_b \to X f_{a'} \dots f_b}\|^2 \equiv \| j_V^a \cdot j^b\|^2 = \sum_{h_a,h_b,h_\ell}\left|j_V^{\mu,h_a h_\ell}(p_1, p_a, p_\ell, p_{\bar{\ell}}) j_\mu^{h_b}(p_{n+2}, p_b)\right|^2,
\end{equation}
with the generalised current~\cite{Andersen:2020yax}
\begin{equation}
  \label{eq:j_V}
  \begin{split}
  j_V^{\mu,h_a h_\ell}(p_a,p_{\ell},p_{\bar{\ell}}, p_1) =&\ \frac{g_V^2}{2}\
     \frac1{p_V^2-M_V^2+i\ \Gamma_V M_V}\ \bar{u}^{h_{\ell}}(p_\ell) \gamma_\alpha
                                               v^{h_{\ell}}(p_{\bar\ell}) \\
& \cdot \left( \frac{ \bar{u}^{h_a}(p_1) \gamma^\alpha (\slashed{p}_V +
  \slashed{p}_1)\gamma^\mu u^{h_a}(p_a)}{(p_V+p_1)^2} +
\frac{ \bar{u}^{h_a}(p_1)\gamma^\mu (\slashed{p}_a - \slashed{p}_V)\gamma^\alpha u^{h_a}(p_a)}{(p_a-p_V)^2} \right).
\end{split}
\end{equation}
$p_V=p_{\ell}+p_{\bar{\ell}}$ is the vector boson momentum, $g_V$ its
coupling to the fermion $f_a$, $M_V$ its mass and $\Gamma_V$ the
width. Note that the emission of a virtual vector boson off $f_a$ implies
that the first $t$-channel momentum is now given by $q_1 = p_a - p_1 -
p_V$.

\subsubsection{Interference}
\label{sec:interference}

So far, we have not considered interference between different
channels. For charged lepton plus neutrino production interference can
arise between emission of the virtual W boson off parton $f_a$ and
emission off parton $f_b$ if both are (anti-)quarks of the same flavour. This contribution
is numerically small as it requires a crossing of the quark line and is
therefore neglected in \HEJ 2.1. Crossing is not required in the
production of a charged lepton-antilepton as the quark line does not change
flavour and same-flavour initial states are no longer required making it
numerically more significant. What is more, there is an additional
interference between production via a photon and production via a Z
boson. Since both effects are non-negligible, we consider the squared LL
matrix element with full interference for this process. Here, we have
to distinguish between $t$-channel momenta $q_{aj}$ for emission of
the vector boson off parton $f_a$ and $q_{bj}$ for emission off
$f_b$. They are given by
\begin{align}
  q_{a1} ={}& p_a - p_1 - p_V, & q_{aj} ={}& q_{a(j-1)} - p_j,\\
  q_{b1} ={}& p_a - p_1, & q_{bj} ={}& q_{b(j-1)} - p_j.
\end{align}
The square of the matrix element then reads~\cite{Andersen:2016vkp}
\begin{equation}
\label{eq:ME_Z}
\begin{split}
\left|\mathcal{M}_{\HEJ}^{f_a f_b\to l\bar{l} f_a \cdot ng \cdot f_b}\right|^2 &=\ (g_s^2)^2 \frac{K_{f_a}K_{f_b}}{4(N_c^2-1)}\ ( g_s^2
C_A)^{n}\  \\  \times \Bigg(& {\frac{\| j^a_{Z\gamma}\cdot
    j^b\|^2}{q^2_{a1}q^2_{a(n+1)}}}
\prod^{n}_{i=1} {\frac{-V^2(q_{ai},
    q_{a(i+1)})}{q^2_{ai} q^2_{a(i+1)}}} \prod_{i=1}^{n+1}
{\exp(\omega^0(q_{ai\perp}) (y_{i+1} - y_i))}\\
+\ &{\frac{\|j^a \cdot j^b_{Z\gamma} \|^2}{q^2_{b1}q^2_{b(n+1)}}}
\prod^{n}_{i=1}{\frac{-V^2(q_{bi}, q_{b(i+1)})}{q^2_{bi} q^2_{b(i+1)}}} \prod_{i=1}^{n+1} {\exp(\omega^0(q_{bi\perp})(y_{i+1} - y_i))} \\
-\ &{\frac{2\Re\{ (j^a_{Z\gamma}\cdot j^b)(\overline{j^a \cdot
      j^b_{Z\gamma}})\}}{\sqrt{q^2_{a1}q^2_{b1}}\sqrt{q^2_{a(n+1)} q^2_{b(n+1)}}}}\\
& \; \times \prod^{n}_{i=1}{\frac{V(q_{ai}, q_{a(i+1)})\cdot V(q_{bi},
    q_{b(i+1)})}{\sqrt{q^2_{ai}q^2_{bi}} \sqrt{q^2_{a(i+1)}q^2_{b(i+1)}}}} \prod_{i=1}^{n+1} {\exp(\omega^0(\sqrt{q_{ai\perp}q_{bi\perp}})(y_{i+1} - y_i))}\Bigg),
\end{split}
\end{equation}
where we have introduced the combined current $j_{Z\gamma} = j_Z +
j_\gamma$ and the notation (cf. equation~(\ref{eq:S_V}))
\begin{equation}
  \label{eq:j_contr}
  (j^a_{V}\cdot j^b)(\overline{j^a \cdot j^b_{V}}) \equiv
  \sum_{h_a,h_b,h_\ell} j_V^{\mu,h_a h_\ell}(p_1, p_a, p_\ell, p_{\bar{\ell}}) j_\mu^{h_b}(p_{n+2}, p_b)  \overline{j_\nu^{h_a}(p_1, p_a) j_V^{\nu,h_b h_\ell}(p_{n+2}, p_b, p_\ell, p_{\bar{\ell}})}.
\end{equation}
The equation~(\ref{eq:ME_Z}) for the square of the matrix element
with interference assumes that both $f_a$ and $f_b$ are
(anti-)quarks. If, for instance, $f_b$ is a gluon instead, terms
involving $j_{Z\gamma}^b$ do not contribute and the expression
simplifies to equation~(\ref{eq:ME_LL}) with the current contraction
in equation~(\ref{eq:S_V}) and $j^a_V \to j^a_{Z\gamma}$.

\subsection{Next-to-leading logarithmic corrections}
\label{sec:NLL_corr}

There are two sources of NLL contributions. First, the matrix elements
for LL configurations, discussed in section~\ref{sec:ME_LL}, receive
NLL corrections. These types of corrections will be considered in
future \HEJ versions. Second, configurations that do not allow the
maximal number of gluonic $t$-channel exchanges according to the
discussion in section~\ref{sec:HEJ2_res} only contribute at higher
logarithmic orders. For instance, this is the case for configurations
where a gluon is produced outside the rapidity order mandated for LL
configurations. Ordering all particles apart from $X$ according to
their rapidity, these processes are denoted by $f_a f_b \to X g
f_{a'}\cdot ng \cdot f_{b'}$, where $f_{a'}$ is an (anti-)quark and
$f_a f_b \to X f_{a'}\cdot ng \cdot f_{b'} g$ with an (anti-)quark
$f_{b'}$. These ``unordered gluon'' NLL configurations were considered
in~\cite{Andersen:2017kfc} and are already
accounted for in \HEJ 2.0.

Studying the production of a W boson together with jets, it is found
that further numerically important NLL configurations arise through
the production of additional pairs of a quark $q$ and an antiquark
$\bar{q}\tinyspace'$. Note that in general the flavours $q$ and $q'$ can be
different if there is a W boson coupling. We distinguish between
``central'' production,
\begin{equation}
  \label{eq:central_qqbar}
  f_a f_b \to X f_{a'}\cdot n_1 g \cdot q \bar{q}\tinyspace' \cdot (n-n_1) g \cdot  f_{b'}, \\
\end{equation}
and ``extremal'' production
\begin{align}
  \label{eq:backward_qqbar}
  g f_b \to{}& X q \bar{q}\tinyspace' \cdot n g \cdot  f_{b'}, \\
  \label{eq:forward_qqbar}
  f_a g \to{}& X f_{a'} \cdot n g \cdot  q \bar{q}\tinyspace'.
\end{align}
In either case, the description is independent of the relative
rapidity ordering between $q$ and $\bar{q}$. In fact, the squares of
the matrix elements for these configurations have the same structure
as at LL accuracy, see equation~(\ref{eq:ME_LL}). Differences only
arise in the Born function ${\cal B}$ and in the
expressions for the $t$-channel momenta $q_i$. \HEJ 2.1 supports
resummation of these configurations for pure multijet production ($X$
is absent) and the production of a leptonically decaying W boson
together with multiple jets ($X = l\bar{\nu}$ or $X = \bar{l}\nu$).

\subsubsection{Pure multijet production}
\label{sec:NLL_jets}

\begin{figure}[tb]
  \centering
  \begin{subfigure}[b]{0.475\textwidth}
  \centering
    \includegraphics{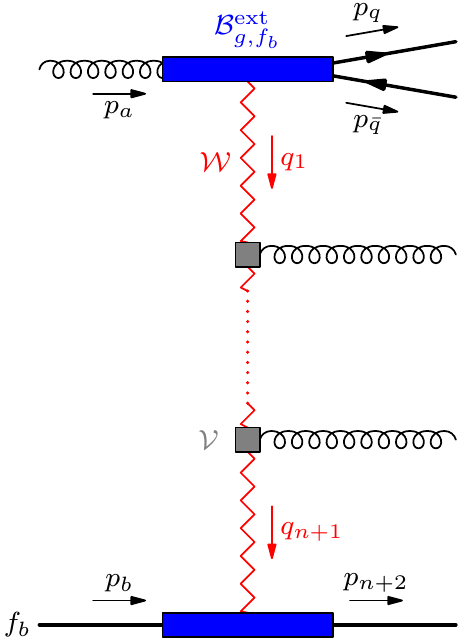}
    \caption{}
  \label{fig:qqbar_ex}
\end{subfigure}
\begin{subfigure}[b]{0.475\textwidth}
  \centering
    \includegraphics{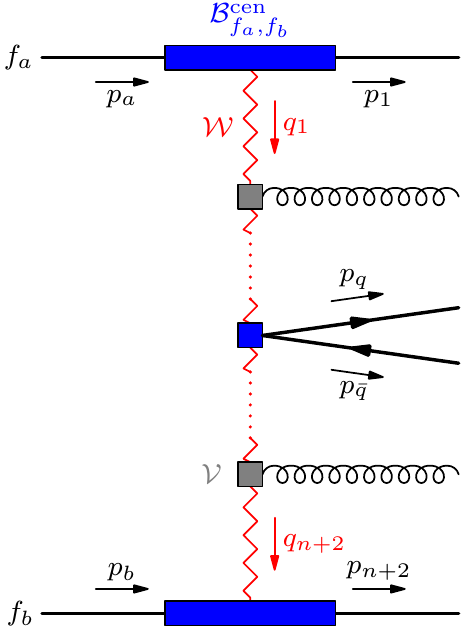}
    \caption{}
    \label{fig:qqbar_cen}
\end{subfigure}
  \caption{Structure of the matrix element for multijet production
involving an extremal (a) or central (b) quark-antiquark pair.}
  \label{fig:qqbar}
\end{figure}

We first consider multijet production with an extremal quark-antiquark
pair. For this, we assume the (same-flavour) quark-antiquark pair to
be emitted in the backward direction; forward emission is completely
analogous. The configurations in question are of the form
\begin{equation}
  \label{eq:backward_qqbar_jets}
  g f_b \to q \bar{q} \cdot n g \cdot  f_{b}.
\end{equation}
In the \HIGHEJ formalism the production of the $q\bar{q}$ pair from
an incoming gluon is described by an effective current
$j_{q\bar{q}}^{\mu,d,h_ah_qh_{\bar{q}}}(p_a, p_q, p_{\bar{q}})$, where
$d$ is a colour index in the adjoint representation, $h_a$ the
helicity of the incoming gluon, and $p_q, h_q$ ($p_{\bar{q}}, h_{\bar{q}}$)
the momentum and helicity of the outgoing (anti-)quark. An explicit
expression for $j_{q\bar{q}}^{\mu,d,h_ah_qh_{\bar{q}}}$ is derived
in~\cite{Andersen:2020yax}. The structure of the matrix element is
illustrated in figure~\ref{fig:qqbar_ex}. Note that due to the emission
of the $q\bar{q}$ pair the first $t$-channel momentum is now given by
$q_1 = p_a - p_q - p_{\bar{q}}$. In summary, the Born function
contributing to the matrix element with extremal $q\bar{q}$ production
reads
\begin{equation}
  \label{eq:B_ex_qqbar}
  {\cal B}^{\text{ext}}_{g,f_b} = (g_s^2)^3 \frac{K_{f_b}}{4(N_C^2-1)} \|S_{gf_b \to q \bar{q} \cdots f_{b}}\|^2 \frac{1}{q_1^2 q_{n+1}^2}\,,
\end{equation}
where
\begin{equation}
  \label{eq:S_ex_qqbar}
\| S_{g f_b \to q \bar{q} \cdots f_{b}}\|^2 \equiv \|j_{q\bar{q}} \cdot j^b\|^2 =  \sum_{h_a,h_b,h_q,h_{\bar{q}}} \left|j_{q\bar{q}}^{\mu,d,h_ah_qh_{\bar{q}}}(p_a, p_q, p_{\bar{q}}) j_\mu^{h_b}(p_b, p_{n_2}) T^d_{b(n+2)}\right|^2.
\end{equation}
Central $q\bar{q}$ production is depicted in figure~\ref{fig:qqbar_cen}
and corresponds to the configuration
\begin{equation}
  \label{eq:central_qqbar_jets}
  f_a f_b \to f_a\cdot n_1 g \cdot q \bar{q} \cdot (n-n_1) g \cdot  f_b. \\
\end{equation}
The central production is described by an effective vertex
$X_{\text{cen}}^{\mu\nu}$ that can be absorbed into the current
contraction. The Born function then reads
\begin{equation}
  \label{eq:B_cen_qqbar}
  {\cal B}^{\text{cen}}_{f_a,f_b} = (g_s^2)^4 \frac{K_{f_a} K_{f_b}}{4(N_C^2-1)} \|S_{f_a f_b \to f_a \cdots q \bar{q} \cdots f_{b}}\|^2 \frac{1}{q_1^2 q_{n+2}^2}
\end{equation}
with the current contraction
\begin{equation}
  \label{eq:S_cen_qqbar}
  \|S_{f_a f_b \to f_a \cdots q \bar{q} \cdots f_{b}}\|^2 \equiv \|j^a \cdot X_{\text{cen}} \cdot j^b \|^2 = \sum_{h_a, h_b}|j_\mu^{h_a}(p_1, p_a) X_{\text{cen}}^{\mu\nu} j_\nu^{h_b}(p_{n+2}, p_b)|^2,
\end{equation}
where $X_{\text{cen}}^{\mu\nu}$ is derived in~\cite{Andersen:2020yax}. The $t$-channel momenta are given by
\begin{equation}
  \label{eq:t_channel_central_qqbar}
 q_i =
 \begin{cases}
   p_a - \sum_{j=1}^i p_i & i \leq n_1 + 1\\
   p_a - p_q - p_{\bar{q}} - \sum_{j=1}^i p_i & i > n_1 + 1
 \end{cases}.
\end{equation}

\subsubsection{W + jets production}
\label{sec:NLL_Wjets}

If an additional leptonically decaying W boson is produced we have to
distinguish between two cases. Similarly to the LL configurations, the W boson may couple to a fermion line associated with an incoming quark or antiquark. This causes a flavour change $f_a \to f_{a'}$ or $f_b \to f_{b'}$. Assuming without loss of generality the latter possibility the corresponding configurations read
\begin{align}
  \label{eq:W_extremal_qqbar}
  gf_b \to{}& X q \bar{q} \cdot n g \cdot f_{b'}  & &(\text{extremal }q \bar{q}),\\
  \label{eq:W_central_qqbar}
  f_a f_b \to{}& X f_{a}\cdot n_1 g \cdot q \bar{q} \cdot (n-n_1) g \cdot  f_{b'} & &(\text{central }q \bar{q}),
\end{align}
where $X = l \bar{\nu}_l$ or $X = \bar{l} \nu_l$.

The Born functions for both configurations are obtained from the
corresponding pure-jets Born functions in
equations~(\ref{eq:B_ex_qqbar}) and (\ref{eq:B_cen_qqbar}) by
replacing the standard current $j^\mu(p_b, p_{n+2})$ with the
generalisation $j_V^\mu(p_b, p_\ell, p_{\bar{\ell}}, p_{n+2})$ for the
emission of a vector boson $V=W$, see equation~(\ref{eq:j_V}).

The second possibility is a coupling of the W boson to the fermion
line of the produced $q\bar{q}\tinyspace'$, where now $q \ne q'$. For
extremal $q\bar{q}\tinyspace'$ production, the Born function is
obtained from equations~(\ref{eq:B_ex_qqbar}) and
(\ref{eq:S_ex_qqbar}) by replacing the current $j_{q\bar{q}}$ with a
new effective current $j_{Wq\bar{q}'}$. In complete analogy the
central production is described by replacing $X_{\text{cen}}$ with a
new effective vertex $X_{W\text{cen}}$ in
equations~(\ref{eq:B_cen_qqbar}), (\ref{eq:S_cen_qqbar}). Explicit
results for $j_{Wq\bar{q}'}$ and $X_{W\text{cen}}$ were obtained
in~\cite{Andersen:2020yax}.


\section{Application: W + jets production }
\label{sec:application}

In order to demonstrate the recent additions to \HEJ we now show in
detail how to obtain a prediction for the production of a W boson with
multiple jets. Specifically, we consider the process $pp \to (W^- \to
e \bar{\nu}_e) + \geq 2$ jets. The parameters are listed in
table~\ref{tab:gen_param}.
\begin{table}[htb]
  \centering
  \renewcommand{\arraystretch}{1.1}
  \begin{tabular}{lp{5mm}l}
    \toprule
    Collider energy && $\sqrt{s} = 13$\,TeV\\[.3em]
    Scales && $\mu_r = \frac{H_T}{2}$\\[.1em]
                    && $\mu_f = \frac{H_T}{2}$\\[.3em]
    PDF set && CT18NLO\\[.3em]
    Electroweak input parameters && $G_F = 1.3663787 \cdot 10^{-5}\,\text{GeV}^{-2}$\\
                    && $m_W = 80.385$\,GeV\\
                    && $\Gamma_W = 2.085$\,GeV\\
                    && $m_Z = 91.187$\,GeV\\
                    && $\Gamma_Z = 2.495$\,GeV\\[.3em]
    Jet definition && anti-$k_t$~\cite{Cacciari:2008gp} \\
                    && $R = 0.7$ \\
                    && $p_\perp > 20$\,GeV\\
    \bottomrule
  \end{tabular}
  \caption{Parameters used for the production of a leptonically
decaying W boson with multiple jets.}
  \label{tab:gen_param}
\end{table}

\subsection{Fixed-order input}
\label{sec:fo_input}

In order to run $\HEJ$, we have to generate leading-order input
events. Any generator producing event files in the Les Houches Event
File (LHEF)~\cite{Alwall:2006yp} format can be used to this end. Here,
we use \texttt{Sherpa}~\cite{Bothmann:2019yzt} with the following run
card. The entries are explained in detail in the \texttt{Sherpa}
documentation.
\begin{lstlisting}[language=sherpa,title=\texttt{\lstname}]
(run){
  EVENTS 10000

  EVENT_OUTPUT LHEF[events_W2j]

  # collider beam
  BEAM_1  2212
  BEAM_ENERGY_1 6500
  BEAM_2  2212
  BEAM_ENERGY_2 6500

  SCALES VAR{H_T2/4}

  # PDF
  PDF_LIBRARY LHAPDFSherpa
  PDF_SET CT18NLO

  MODEL SM

  # electroweak parameters
  EW_SCHEME 3
  GF 1.1663787e-05

  MASS[24] 80.385
  WIDTH[24] 2.085

  MASS[23] 91.187
  WIDTH[23] 2.495

  # massless charm quark
  MASSIVE[4] 0
  YUKAWA[4] 0.
  MASS[4] 0.

  # massless bottom quark
  MASSIVE[5] 0
  YUKAWA[5] 0.
  MASS[5] 0.

  ME_SIGNAL_GENERATOR Comix
  EVENT_GENERATION_MODE Weighted

  # disable everything beyond fixed order
  FRAGMENTATION Off
  YFS_MODE 0
  MI_HANDLER None
  SHOWER_GENERATOR None
  CSS_MAXEM 0
  BEAM_REMNANTS 0
}(run)

(processes){
  Process 93 93 -> 24[a] 93 93
  Decay 24[a] -> -11 12
  Order (*,2)
  End process
}(processes)

(selector){
  # require 2 anti-kt jets with pt > 18 GeV and R = 0.7
  FastjetFinder  antikt 2  18.  0.0  0.7
}(selector)
\end{lstlisting}
Note that we have chosen a minimum jet transverse momentum of
$18$\,GeV instead of the $20$\,GeV listed in
table~\ref{tab:gen_param}. The reason for this is that the \HEJ
resummation can change the transverse momenta of the jets slightly
compared to the input leading-order events. It is therefore
recommended to choose a minimum transverse momentum that is at least
$10\%$ smaller in the fixed-order generation. It is also prudent to
ensure that the final predictions do not change when choosing an even
smaller value.

The contribution from events containing such ``soft'' jets with
transverse momenta below the value required in the final analysis is
numerically small. This means that a significant amount of computing
time can be saved by generating two separate input event samples: one
sample with low statistics in which each event contains at least one
soft jet, and one large sample without any soft jets. We then apply
resummation separately to each of the samples. Since the two samples
cover distinct regions of the fixed-order phase space, the final
prediction is simply the set of all generated events. For simplicity,
we do not consider this optimisation in the following.

With the above \texttt{Run.dat} run card, an event file
\texttt{events\_W2j.lhe} can be generated by running
\begin{lstlisting}[language=sh]
Sherpa
\end{lstlisting}
in the same directory. The generated events will contain exactly two
jets, but it is straightforward to generate higher-multiplicity
samples by increasing the required number of jets in \lstinline{FastjetFinder},
adding the corresponding number of \lstinline{93} entries to the
\lstinline{Process} final state, and adjusting the name of the output
file in the \lstinline{EVENT_OUTPUT}. \HEJ can then be used to merge
samples with different multiplicities. Typically, the contribution of
higher jet multiplicities in the analysis will be decreasing, and it
is often enough to consider at most four or five jets. One can save
computing time at the price of sacrificing leading-order accuracy by
using the \HEJ fixed-order generator (\HEJFOG) for high
multiplicities. An example is shown in appendix~\ref{sec:Wjets_HEJFOG}.

\subsection{Resummation with \HEJ}
\label{sec:HEJ_run}

In addition to the fixed-order input, \HEJ also requires a
configuration file. A template \texttt{config.yml} is included in the
\HEJ source code. Adopting the current set of parameters from
table~\ref{tab:gen_param} and enabling resummation for all supported
NLL configurations we get
\begin{lstlisting}[language=yaml,title=\texttt{\lstname}]
## Number of attempted resummation phase space points for each input event
trials: 10

resummation jets:       # resummation jet properties
  algorithm: antikt     # jet clustering algorithm
  R: 0.7                # jet R parameter
  min pt: 20            # minimum jet transverse momentum

fixed order jets:       # properties of input jets
  min pt: 18
  # by default, algorithm and R are like for resummation jets

## Treatment of the various event classes
## the supported settings are: reweight, keep, discard
## non-resummable events cannot be reweighted
event treatment:
  FKL: reweight
  # Enable resummation for NLL configurations
  unordered: reweight
  extremal qqbar: reweight
  central qqbar: reweight
  non-resummable: keep

## Central scale choice or choices
scales: H_T/2

## Selection of random number generator and seed
random generator:
  name: mixmax
  seed: 1

## Whether or not to include higher order logs
log correction: false

## Vacuum expectation value
vev: 246.2196508

## Properties of the weak gauge bosons
particle properties:
  W:
    mass: 80.385
    width: 2.085
  Z:
    mass: 91.187
    width: 2.495
  Higgs:
    mass: 125
    width: 0.004165
\end{lstlisting}
We can now generate \HEJ events and calculate the total resummed cross section.
If \HEJ is not installed, it can still be run using the Docker
virtualisation software~\cite{docker} and the following command.
\begin{lstlisting}[language=sh]
docker run -v $PWD:$PWD -w $PWD hejdock/hej HEJ config_Wjets.yml events_W2j.lhe
\end{lstlisting}
If a \HEJ installation is available, we can instead run
\begin{lstlisting}[language=sh]
HEJ config_Wjets.yml events_W2j.lhe
\end{lstlisting}

\begin{figure}[htb]
  \centering
  \includegraphics[width=0.47\linewidth]{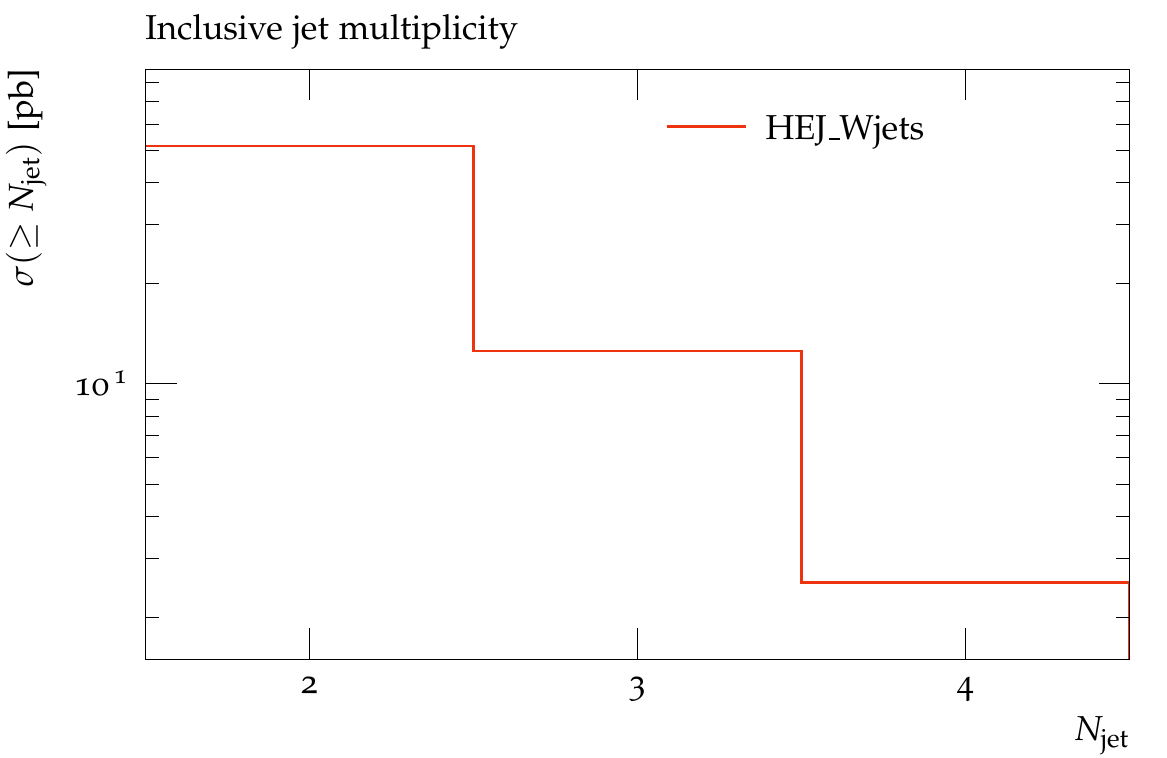}
  \includegraphics[width=0.47\linewidth]{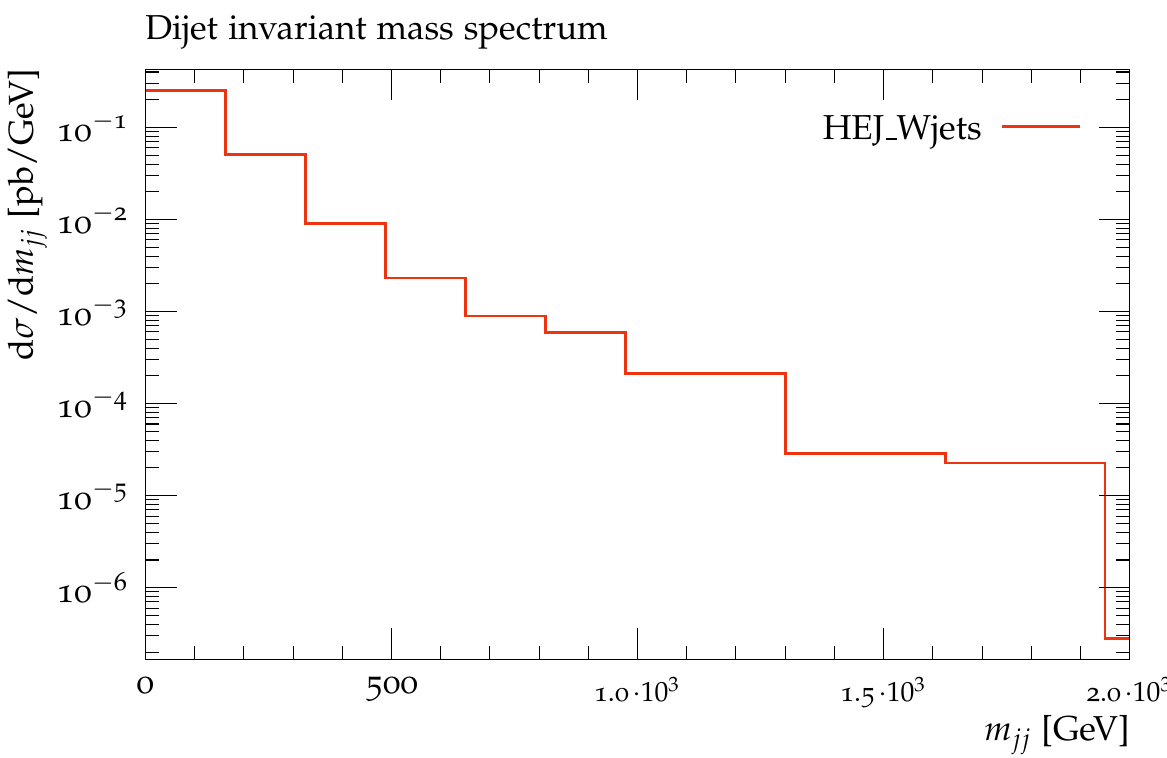}
  \caption{Inclusive $N$-jet cross sections (left) and dijet invariant
    mass spectrum (right) obtained with \texttt{Sherpa} and \HEJ 2.1 for
    the production of a W boson with at least two jets.}
  \label{fig:distr}
\end{figure}

If \HEJ was compiled with support for
\texttt{Rivet}~\cite{Bierlich:2019rhm},\footnote{The \HEJ Docker container includes all optional dependencies.} the generated events can be
forwarded directly to the \texttt{MC\_WJETS} analysis by adding the
following lines to \texttt{config\_Wjets.yml} before running \HEJ:
\begin{lstlisting}
analyses:
  - rivet: MC_WJETS
    output: HEJ_W2j
\end{lstlisting}
Another possibility is to create an output event file for manual
analysis. After compiling \HEJ with support for the \texttt{HepMC
2}~\cite{Dobbs:2001ck} format, we can add the following entry to
\texttt{config\_Wjets.yml}:
\begin{lstlisting}
event output:
  - HEJ_W2j.hepmc2
\end{lstlisting}
We can then run \HEJ as before and feed the generated events into the
\texttt{MC\_WJETS} analysis with
\begin{lstlisting}[language=sh]
rivet -a MC_WJETS -o HEJ_W2j.yoda HEJ_W2j.hepmc2
\end{lstlisting}
Results for higher jet multiplicities are obtained in the same way
after adjusting the file names in the configuration files and run
commands. To guarantee statistical independence it is also recommended
to change the \lstinline{seed} in \texttt{config\_Wjets.yml}. To
obtain a prediction that is inclusive in jet multiplicity, we combine
the \texttt{Rivet} output with\footnote{In older \texttt{Rivet}
versions \lstinline{yodamerge --add} can be used instead of
\lstinline{yodastack}.}
\begin{lstlisting}[language=sh]
yodastack -o HEJ_Wjets.yoda HEJ_W*j.yoda
\end{lstlisting}
Finally,
\begin{lstlisting}[language=sh]
rivet-mkhtml HEJ_Wjets.yoda
\end{lstlisting}
produces analysis plots. As examples, we show the inclusive $N$-jet
cross sections and the invariant mass distribution of the two hardest
jets obtained from fixed-order input events with up to four jets in figure~\ref{fig:distr}.


\section{Conclusions}
\label{sec:conclusions}

This article accompanies the release of the event generator, \HEJ~\texttt{2.1}
which provides predictions for hadronic colliders which contain contributions at
all orders in $\alpha_s$ to achieve
leading-logarithmic accuracy in $\hat s/p_\perp^2$.  The output is given as
exclusive events meaning that predictions can be obtained for arbitrary
experimental cuts and analyses.

Here, after a brief outline of the general \HEJ formalism, we have described the new
extensions which have broadened the scope of the all-order predictions.
Specifically, new matrix elements which allow for interference between channels
with \emph{different} assignments of effective $t$-channel momenta have been
developed to give an accurate description of $pp\to (Z/\gamma^*\to \ell
\bar\ell) +\ge2$~jets (section~\ref{sec:interference}).  Secondly, the full set of NLL
contributions corresponding to processes which have no LL component have been
incorporated through the derivation of necessary additional Born processes
(section \ref{sec:NLL_corr}).  These form a well-defined, gauge-invariant subset
of the full NLL corrections.

Finally, we have provided a worked example explicitly showing all the steps
necessary to generate predictions for $pp\to (W^-\to e\bar\nu_e) +\ge2$~jets.

\section*{Acknowledgements}
\label{sec:acknowledgements}

\HEJ uses \texttt{FastJet}~\cite{Cacciari:2011ma}, \texttt{LHAPDF}~\cite{Buckley:2014ana}, and
\texttt{FORM}~\cite{Vermaseren:2000nd}.

We are pleased to acknowledge funding from the UK
Science and Technology Facilities Council, the Royal Society, the ERC
Starting Grant 715049 ``QCDforfuture'', the Marie Sk{\l}odowska-Curie
Innovative Training Network MCnetITN3 (grant agreement no.~722104),
and the EU TMR network SAGEX agreement No.  764850 (Marie
Skłodowska-Curie).


\appendix

\section{W + jets production with the \HEJ fixed-order generator}
\label{sec:Wjets_HEJFOG}

The generation of fixed-order events with high jet multiplicities can
become prohibitively expensive in computing time. In such cases the
\HEJ fixed-order generator (\HEJFOG) can be used to quickly obtain
results in the high-energy approximation. A typical use case is to
produce \HEJ input files with exact leading-order accuracy up to a
given jet multiplicity as discussed in section~\ref{sec:fo_input} and
supplement them with high-multiplicity event files produced with the
\HEJFOG.

As an example, we slightly modify the
configuration file \texttt{configFO.yml} included in the \HEJ source
code to produce events for the process $pp \to (W^- \to e \bar{\nu}_e)
+ 5$ jets.
\begin{lstlisting}[language=yaml,title=\texttt{\lstname}]
## Number of generated events
events: 10000

jets:
  min pt: 18        # minimal jet pt, should be slightly below analysis cut
  peak pt: 20       # peak pt of jets, should be at analysis cut
  algorithm: antikt # jet clustering algorithm
  R: 0.7            # jet R parameter
  max rapidity: 5   # maximum jet rapidity

## Particle beam
beam:
  energy: 6500
  particles: [p, p]

## PDF ID for CT18NLO
pdf: 14400

## Scattering process
process: p p => Wm 5j

## Particle decays (multiple decays are allowed)
decays:
  Wm: {into: [e-, nu_e_bar]}

## Fraction of events with two extremal emissions in one direction
## that contain an subleading emission e.g. unordered emission
subleading fraction: 0.05

## Allow different subleading configurations.
## By default all processes are allowed.
## This does not check if the processes are implemented in HEJ!
#
subleading channels:
  - unordered
  - central qqbar
  - extremal qqbar

## Central scale choice
scales: H_T/2

## Event output files
event output:
  - events_W5j.lhe

## Selection of random number generator and seed
random generator:
  name: mixmax
  seed: 5

## Vacuum expectation value
vev: 246.2196508

## Properties of the weak gauge bosons
particle properties:
  W:
    mass: 80.385
    width: 2.085
  Z:
    mass: 91.187
    width: 2.495
  Higgs:
    mass: 125
    width: 0.004165

unweight:
  sample size: 10000
  max deviation: 0
\end{lstlisting}
To generate the event file \texttt{events\_W5j.lhe} run
\begin{lstlisting}
docker run -v $PWD:$PWD -w $PWD hejdock/hej HEJFOG config_HEJFOG.yml
\end{lstlisting}
when using the \HEJ Docker container or
\begin{lstlisting}
HEJFOG config_HEJFOG.yml
\end{lstlisting}
when using a local \HEJ installation.


\bibliographystyle{JHEP}
\bibliography{papers}

\providecommand{\href}[2]{#2}\begingroup\raggedright\begin{thebibliography}{10}

\bibitem{Andersen:2009nu}
J.~R. Andersen and J.~M. Smillie, \emph{{Constructing All-Order Corrections to
  Multi-Jet Rates}},
  \href{http://dx.doi.org/10.1007/JHEP01(2010)039}{\emph{JHEP} {\bf 1001}
  (2010) 039}, [\href{http://arxiv.org/abs/0908.2786}{{\tt 0908.2786}}].

\bibitem{Andersen:2009he}
J.~R. Andersen and J.~M. Smillie, \emph{{The Factorisation of the t-channel
  Pole in Quark-Gluon Scattering}},
  \href{http://dx.doi.org/10.1103/PhysRevD.81.114021}{\emph{Phys.Rev.} {\bf
  D81} (2010) 114021}, [\href{http://arxiv.org/abs/0910.5113}{{\tt
  0910.5113}}].

\bibitem{Andersen:2011hs}
J.~R. Andersen and J.~M. Smillie, \emph{{Multiple Jets at the LHC with High
  Energy Jets}}, \href{http://dx.doi.org/10.1007/JHEP06(2011)010}{\emph{JHEP}
  {\bf 1106} (2011) 010}, [\href{http://arxiv.org/abs/1101.5394}{{\tt
  1101.5394}}].

\bibitem{Andersen:2017kfc}
J.~R. Andersen, T.~Hapola, A.~Maier and J.~M. Smillie, \emph{{Higgs Boson Plus
  Dijets: Higher Order Corrections}},
  \href{http://dx.doi.org/10.1007/JHEP09(2017)065}{\emph{JHEP} {\bf 09} (2017)
  065}, [\href{http://arxiv.org/abs/1706.01002}{{\tt 1706.01002}}].

\bibitem{Andersen:2020yax}
J.~R. Andersen, J.~A. Black, H.~M. Brooks, E.~P. Byrne, A.~Maier and J.~M.
  Smillie, \emph{{Combined subleading high-energy logarithms and NLO accuracy
  for W production in association with multiple jets}},
  \href{http://dx.doi.org/10.1007/JHEP04(2021)105}{\emph{JHEP} {\bf 04} (2021)
  105}, [\href{http://arxiv.org/abs/2012.10310}{{\tt 2012.10310}}].

\bibitem{ATLAS:2011yyh}
{\scshape ATLAS} collaboration, G.~Aad et~al., \emph{{Measurement of dijet
  production with a veto on additional central jet activity in $pp$ collisions
  at $\sqrt{s}=7$ TeV using the ATLAS detector}},
  \href{http://dx.doi.org/10.1007/JHEP09(2011)053}{\emph{JHEP} {\bf 09} (2011)
  053}, [\href{http://arxiv.org/abs/1107.1641}{{\tt 1107.1641}}].

\bibitem{CMS:2012rfo}
{\scshape CMS} collaboration, S.~Chatrchyan et~al., \emph{{Measurement of the
  inclusive production cross sections for forward jets and for dijet events
  with one forward and one central jet in $pp$ collisions at $\sqrt{s}=7$
  TeV}}, \href{http://dx.doi.org/10.1007/JHEP06(2012)036}{\emph{JHEP} {\bf 06}
  (2012) 036}, [\href{http://arxiv.org/abs/1202.0704}{{\tt 1202.0704}}].

\bibitem{CMS:2012xfg}
{\scshape CMS} collaboration, S.~Chatrchyan et~al., \emph{{Ratios of dijet
  production cross sections as a function of the absolute difference in
  rapidity between jets in proton-proton collisions at $\sqrt{s}=7$ TeV}},
  \href{http://dx.doi.org/10.1140/epjc/s10052-012-2216-6}{\emph{Eur. Phys. J.
  C} {\bf 72} (2012) 2216}, [\href{http://arxiv.org/abs/1204.0696}{{\tt
  1204.0696}}].

\bibitem{ATLAS:2014lzu}
{\scshape ATLAS} collaboration, G.~Aad et~al., \emph{{Measurements of jet
  vetoes and azimuthal decorrelations in dijet events produced in $pp$
  collisions at $\sqrt{s}=7\,\mathrm{TeV}$ using the ATLAS detector}},
  \href{http://dx.doi.org/10.1140/epjc/s10052-014-3117-7}{\emph{Eur. Phys. J.
  C} {\bf 74} (2014) 3117}, [\href{http://arxiv.org/abs/1407.5756}{{\tt
  1407.5756}}].

\bibitem{Andersen:2012gk}
J.~R. Andersen, T.~Hapola and J.~M. Smillie, \emph{{W Plus Multiple Jets at the
  LHC with High Energy Jets}},
  \href{http://dx.doi.org/10.1007/JHEP09(2012)047}{\emph{JHEP} {\bf 1209}
  (2012) 047}, [\href{http://arxiv.org/abs/1206.6763}{{\tt 1206.6763}}].

\bibitem{D0:2013gro}
{\scshape D0} collaboration, V.~M. Abazov et~al., \emph{{Studies of W boson
  plus jets production in $p\bar{p}$ collisions at $\sqrt{s}=1.96$ TeV}},
  \href{http://dx.doi.org/10.1103/PhysRevD.88.092001}{\emph{Phys. Rev. D} {\bf
  88} (2013) 092001}, [\href{http://arxiv.org/abs/1302.6508}{{\tt 1302.6508}}].

\bibitem{ATLAS:2014fjg}
{\scshape ATLAS} collaboration, G.~Aad et~al., \emph{{Measurements of the W
  production cross sections in association with jets with the ATLAS detector}},
  \href{http://dx.doi.org/10.1140/epjc/s10052-015-3262-7}{\emph{Eur. Phys. J.
  C} {\bf 75} (2015) 82}, [\href{http://arxiv.org/abs/1409.8639}{{\tt
  1409.8639}}].

\bibitem{Andersen:2016vkp}
J.~R. Andersen, J.~J. Medley and J.~M. Smillie, \emph{{$Z/\gamma^{*}$ plus
  multiple hard jets in high energy collisions}},
  \href{http://dx.doi.org/10.1007/JHEP05(2016)136}{\emph{JHEP} {\bf 05} (2016)
  136}, [\href{http://arxiv.org/abs/1603.05460}{{\tt 1603.05460}}].

\bibitem{Andersen:2008gc}
J.~R. Andersen, V.~Del~Duca and C.~D. White, \emph{{Higgs Boson Production in
  Association with Multiple Hard Jets}},
  \href{http://dx.doi.org/10.1088/1126-6708/2009/02/015}{\emph{JHEP} {\bf 02}
  (2009) 015}, [\href{http://arxiv.org/abs/0808.3696}{{\tt 0808.3696}}].

\bibitem{Andersen:2018tnm}
J.~R. Andersen, T.~Hapola, M.~Heil, A.~Maier and J.~M. Smillie,
  \emph{{Higgs-boson plus Dijets: Higher-Order Matching for High-Energy
  Predictions}}, \href{http://dx.doi.org/10.1007/JHEP08(2018)090}{\emph{JHEP}
  {\bf 08} (2018) 090}, [\href{http://arxiv.org/abs/1805.04446}{{\tt
  1805.04446}}].

\bibitem{Andersen:2018kjg}
J.~R. Andersen, J.~D. Cockburn, M.~Heil, A.~Maier and J.~M. Smillie,
  \emph{{Finite Quark-Mass Effects in Higgs Boson Production with Dijets at
  Large Energies}},
  \href{http://dx.doi.org/10.1007/JHEP04(2019)127}{\emph{JHEP} {\bf 04} (2019)
  127}, [\href{http://arxiv.org/abs/1812.08072}{{\tt 1812.08072}}].

\bibitem{Andersen:2019yzo}
J.~R. Andersen, T.~Hapola, M.~Heil, A.~Maier and J.~Smillie, \emph{{HEJ 2: High
  Energy Resummation for Hadron Colliders}},
  \href{http://dx.doi.org/10.1016/j.cpc.2019.06.022}{\emph{Comput.Phys.Commun.}
  {\bf 245} (2019) }, [\href{http://arxiv.org/abs/1902.08430}{{\tt
  1902.08430}}].

\bibitem{Lipatov:1976zz}
L.~N. Lipatov, \emph{{Reggeization of the Vector Meson and the Vacuum
  Singularity in Nonabelian Gauge Theories}}, {\emph{Sov. J. Nucl. Phys.} {\bf
  23} (1976) 338--345}.

\bibitem{Cacciari:2008gp}
M.~Cacciari, G.~P. Salam and G.~Soyez, \emph{{The anti-$k_t$ jet clustering
  algorithm}},
  \href{http://dx.doi.org/10.1088/1126-6708/2008/04/063}{\emph{JHEP} {\bf 04}
  (2008) 063}, [\href{http://arxiv.org/abs/0802.1189}{{\tt 0802.1189}}].

\bibitem{Alwall:2006yp}
J.~Alwall et~al., \emph{{A Standard format for Les Houches event files}},
  \href{http://dx.doi.org/10.1016/j.cpc.2006.11.010}{\emph{Comput. Phys.
  Commun.} {\bf 176} (2007) 300--304},
  [\href{http://arxiv.org/abs/hep-ph/0609017}{{\tt hep-ph/0609017}}].

\bibitem{Bothmann:2019yzt}
{\scshape Sherpa} collaboration, E.~Bothmann et~al., \emph{{Event Generation
  with Sherpa 2.2}},
  \href{http://dx.doi.org/10.21468/SciPostPhys.7.3.034}{\emph{SciPost Phys.}
  {\bf 7} (2019) 034}, [\href{http://arxiv.org/abs/1905.09127}{{\tt
  1905.09127}}].

\bibitem{docker}
``Docker.''

\bibitem{Bierlich:2019rhm}
C.~Bierlich et~al., \emph{{Robust Independent Validation of Experiment and
  Theory: Rivet version 3}},
  \href{http://dx.doi.org/10.21468/SciPostPhys.8.2.026}{\emph{SciPost Phys.}
  {\bf 8} (2020) 026}, [\href{http://arxiv.org/abs/1912.05451}{{\tt
  1912.05451}}].

\bibitem{Dobbs:2001ck}
M.~Dobbs and J.~B. Hansen, \emph{{The HepMC C++ Monte Carlo event record for
  High Energy Physics}},
  \href{http://dx.doi.org/10.1016/S0010-4655(00)00189-2}{\emph{Comput. Phys.
  Commun.} {\bf 134} (2001) 41--46}.

\bibitem{Cacciari:2011ma}
M.~Cacciari, G.~P. Salam and G.~Soyez, \emph{{FastJet User Manual}},
  \href{http://dx.doi.org/10.1140/epjc/s10052-012-1896-2}{\emph{Eur. Phys. J.
  C} {\bf 72} (2012) 1896}, [\href{http://arxiv.org/abs/1111.6097}{{\tt
  1111.6097}}].

\bibitem{Buckley:2014ana}
A.~Buckley, J.~Ferrando, S.~Lloyd, K.~Nordstr\"om, B.~Page, M.~R\"ufenacht
  et~al., \emph{{LHAPDF6: parton density access in the LHC precision era}},
  \href{http://dx.doi.org/10.1140/epjc/s10052-015-3318-8}{\emph{Eur. Phys. J.
  C} {\bf 75} (2015) 132}, [\href{http://arxiv.org/abs/1412.7420}{{\tt
  1412.7420}}].

\bibitem{Vermaseren:2000nd}
J.~A.~M. Vermaseren, \emph{{New features of FORM}},
  \href{http://arxiv.org/abs/math-ph/0010025}{{\tt math-ph/0010025}}.

\end{thebibliography}\endgroup

\end{document}